\documentclass[aps,prl,twocolumn,superscriptaddress]{revtex4}
\usepackage{amssymb}
\usepackage{amsmath}
\usepackage{graphicx}

\begin{document}

\title{Comment on "Consistent Interpretation of the Low-Temperature
Magnetotransport in Graphite Using the Slonczewski-Weiss-McClure 3D
Band-Structure Calculations" }
\author{Igor A. Luk'yanchuk}
\affiliation{University of Picardie Jules Verne, Laboratory of Condensed Matter Physics,
Amiens, 80039, France}
\author{Yakov Kopelevich}
\affiliation{Instituto de F\'{\i}sica "Gleb Wataghin", Universidade Estadual de Campinas,
UNICAMP 13083-970, Campinas, Sao Paulo, Brazil}
\date{\today}
\maketitle

In \cite{Lukyanchuk2004,Lukyanchuk2006} we have shown that substantial part
of conductivity in graphite is provided by holes (\emph{h}) with massless
linear spectrum $\varepsilon (p)=v|{p_{\perp }|}$ - Dirac Fermions (DF) that
coexist with massive normal carriers (NC) - electrons (\emph{e}) with $%
\varepsilon (p)={p^{2}}/{2m}^{\ast }$. Existence of such quantity of DF does
not follow from the classical Slonczewski Weiss and McClure (SWM) band model
and can signify that at least part of carbon layers behaves like independent
grahenes.

In a recent Letter \cite{Schneider2009} Schneider \emph{et al.}
revised our conclusion pointed that both types of carriers are
massive and are described
by SWM model. Since both \cite{Lukyanchuk2004,Lukyanchuk2006} and \cite%
{Schneider2009} use the same method of phase determination of Shubnikov de
Haas (SdH) oscillation we comment here that the controversy originates from
the improper treatment of experimental results in \cite{Schneider2009}.

The sense of the method is to extract the phase $\phi_1$ from the quantum
oscillation of conductivity:
\begin{equation}
\sigma _{xx}(B)=\sum_{l=1}^\infty a_{l}\cos [2\pi l{\frac{\mu
}{\hbar \omega _{c}}+}\varphi_l l],  \label{Phase}
\end{equation}%
by noting that $\varphi_1 =\pi $ for NC and $0$ for DF ($\mu $ is the
chemical potential, $\hbar \omega _{c}= {\frac{ {eB} }{{m^{\ast }c}}}$ for
NC and ${\frac{ev^{2}B }{c\mu}} $ for DF).

\begin{figure}[ht]
\centering
\includegraphics [width=6.8 cm] {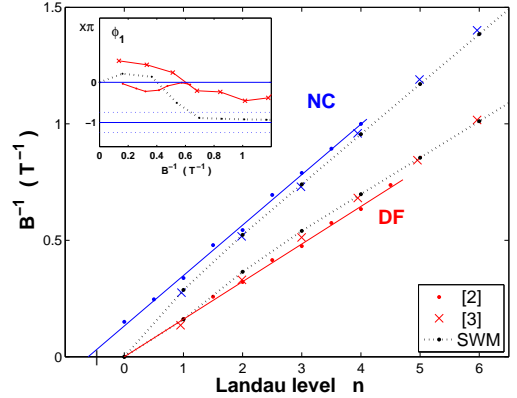}
\caption{Max (min) of SdH oscillation for two groups of carriers as
function of their LL number $n$. Linear extrapolation to $B^{-1}=0$
gives the phase - $\protect\varphi_1=2\protect\pi n_0$. Insert shows
the dependence of differential $\protect\varphi_1$ on $B^{-1}$}
\label{fancompar}
\end{figure}

Note first that presented in Fig.~\ref{fancompar} method to find $\varphi_1$
shows the remarkable coincidence between our \cite{Lukyanchuk2006} and
Schneider \emph{et al.} results. The lower line corresponds to carriers with
higher frequency (HF). From its extrapolation to $B^{-1}=0$ we clearly see
that at $B\rightarrow \infty $ the lowest LL ($n=0$) is placed exactly at $%
E=0$ and that $\varphi=0$, as it follows for DF. Similarly the low frequency
(LF) carriers with $\varphi_1 \sim \pi$ are attributed to NC.

Schneider \emph{et al.} argue that these data can not be used because "in
the quantum limit the Fermi energy [$\mu$ in (\ref{Phase})] is no longer
constant as carriers are transferred between the electron and holes".

To verify this doubt we present in Fig.~\ref{fancompar} the
calculated within SWM model diagram of $B_{n}^{-1}$ at which SdH
oscillation exhibits maxima:
\begin{equation}
B_{n}^{-1}=\sqrt{n(n+1)}\left[ 1-\frac{\Delta \mu _{n}}{\mu _{0}}\right]
B_{0}^{-1}  \label{correct}
\end{equation}%
The first (band) factor \cite{Smrcka2009} generalizes the used in \cite%
{Lukyanchuk2004,Lukyanchuk2006} quasi-classical $n+{\frac{1}{2}}$
quantization. The taken from \cite{Ono1966} correction to $\mu $ is
due to electron-hole cross-talk.

Next, we trace the differential phase $\varphi _{1}(B^{-1})=-2\pi \left[
n(B^{-1})-B^{-1}(dn/dB^{-1})\right] $ for HF carriers (smoothed by 2-point
moving average) and observe that the SWM curve, as was mentioned in \cite%
{Schneider2009}, has the strong nonlinear deviation from $-\pi $ at $B>2T$.
Our data don't demonstrate such non-linearity whereas Schneider' \emph{et al.%
} stay close to $\phi _{1}=0$ and don't drop together with SWM curve to $%
\varphi _{1}=-\pi $ at $2T>B>0.7T$. Contradiction with SWM model and
closeness of $\varphi_1$ to $0$ confirms the existence of DF in graphite.

Note that proposed in \cite{Schneider2009} extrapolation of $\varphi _{1}$
from fields $B<0.7$~T is not reliable. Thus, for the presented in Fig.~{2}e
of \cite{Schneider2009} phase-frequency analysis of HF carriers one gets $%
\varphi _{1}\simeq \left( 0.56\pm 0.6\right) \pi $. This value and
error-bar, determined as FWHM of 2D Gaussian projected on phase-axis
are insufficient to discriminate between the DF and NC.

The work was supported by programs ANR-LoMaCoQuP, FP7-ROBOCON, CNPq and
FAPESP.


\begin{thebibliography}{9}
\bibitem{Lukyanchuk2004} I. A. Luk'yanchuk and Y. Kopelevich, Phys. Rev.
Lett. \textbf{93}, 166402 (2004). 

\bibitem{Lukyanchuk2006} I. A. Luk'yanchuk and Y. Kopelevich, Phys. Rev.
Lett. \textbf{97}, 256801 (2006). 

\bibitem{Schneider2009} J. M. Schneider, M. Orlita, M. Potemski, and D. K.
Maude, Phys. Rev. Lett. \textbf{102}, 166403 (2009).

\bibitem{Smrcka2009} L.Smr{\v{c}}ka, N.A.Goncharuk, Phys.Rev. \textbf{B80},
073403 (2009)

\bibitem{Ono1966} K. Sugihara, S. Ono, J. Phys. Soc. Jpn. \textbf{B} \textbf{%
21}, 631 (1966)
\end{thebibliography}
\end{document}